\documentclass[10pt]{elsart}
\usepackage[dvips]{graphicx}
\usepackage[nomarkers]{endfloat}  
\usepackage{amsmath,amssymb}
\usepackage{url}              
\usepackage{multirow}
\newcommand{\urlBiBTeX}[1]{\url{#1}}
\usepackage{textcomp}  
\journal{arXiv.org (Published, DOI: 10.1016/j.nimb.2006.03.190) }

\begin{document}

\begin{frontmatter}
\title{Simultaneous PIXE and RBS data analysis using Bayesian Inference with the DataFurnace code}

\author[ITN,CMAM]{C. Pascual-Izarra \corauthref{ca}},
\corauth[ca]{Corresponding author: phone +34 91 5616800, fax +34 91 5859413}
\ead{carlos.pascual@iem.cfmac.csic.es}
\ead[url]{http://cpixe.sourceforge.net}
\author[ITN]{M. A. Reis},
\author[ITN]{N. P. Barradas},

\address[ITN]{Instituto Tecnol\'ogico e Nuclear, E.N. 10, 2686-953 Sacav\'em
Codex, Portugal}
\address[CMAM]{CMAM, Universidad Aut\'onoma de Madrid, E-28049 Spain}

\begin{abstract}
The Rutherford Backscattering Spectroscopy (RBS) and Particle Induced X-ray Emission (PIXE) techniques can be used to obtain complementary information about the characteristics of a sample but, traditionally, a gap has separated the available computer codes for analyzing data from each technique, being hard to simultaneously analyze data from the same sample. The recent development of a free and open source library, LibCPIXE, for PIXE simulation and analysis of arbitrary multilayered samples, has permitted to integrate this technique into the DataFurnace code which already handles many other IBA techniques such as Rutherford and non-Rutherford backscattering, elastic recoil detection, and non-resonant nuclear reaction analysis. The fitting capabilities of DataFurnace can therefore now be applied to PIXE spectra as well, including the Bayesian Inference analysis and the simultaneous and coherent fitting of multiple spectra from different techniques. Various examples are presented in which the simultaneous RBS and PIXE analysis allows us to obtain consistent results that cannot be obtained by independent analysis of the data from each technique.
\end{abstract}
\begin{small}
Note: This is a preprint of a work published in Nuclear Instruments and Methods in Physics Research Section B 249, 780-783 (2006). Please cite the published one.
\end{small}
\end{frontmatter}
\section{Introduction}
Despite of the close relation of particle-based IBA techniques (RBS, ERDA,\ldots) with photon-based ones (PIXE, PIGE,\ldots), and despite the fact that they are often used to obtain complementary information about the same sample, the most popular IBA analysis codes are focused either on one or the other category and data cannot be analyzed in a coherent and unified way. An attempt to unify PIXE and RBS analysis had been already done \cite{web:DAN32} by building a common interface to the well known RUMP \cite{web:RUMP} and GUPIX \cite{web:GUPIX} programs. In this work we present a more ambitious initiative in terms of integrated analysis of IBA techniques: the implementation of the support for PIXE in the DataFurnace code (which already supported RBS, ERDA, and non-resonant NRA), achieved by using the recently developed LibCPIXE free library for PIXE yield simulation \cite{LibCPIXE}. This implies that all the fitting capabilities already available in DataFurnace for other techniques can now be applied to PIXE, including simulated annealing fits and the Bayesian inference error estimation.

\section{DataFurnace $+$ LibCPIXE}\label{s:methodology}
The integration of LibCPIXE with DataFurnace, incorporates PIXE to the repertory of supported techniques in the latter code. This opens up the possibility of \emph{simultaneously} analyzing data from several spectra of a given sample and thus merging the information from all of them, producing a self--consistent result. Note that the experimental spectra can be obtained with different techniques (now including PIXE) as well as with the same technique under different experimental conditions in order to obtain complementary information about the sample.

In contrast with the most widely used programs for PIXE data analysis, which make use of nonlinear least squares or Marquardt fitting algorithms \cite{IAEA-PIXE}, the DataFurnace code implements simulated annealing and Bayesian inference algorithms. The combination of these algorithms has proven to be a very powerful tool when dealing with IBA problems in which one may have multiple and/or suboptimal formal solutions \cite{Barradas03-BI-IBA}. In relation with PIXE analysis, this provides a very flexible fitting procedure that, \emph{e.g}, does not pose any limit to the analysis of layered samples with a given element present in more than one layer (which is, currently, a limitation in the GUPIX code \cite{GUPIX95}). The Bayesian inference has also the advantage of providing a mathematically rigorous estimation of the errors in the fitted parameters (concentration profiles for each element in the sample).

The treatment of the PIXE data in DataFurnace differs from that done for other techniques or that done by other PIXE analysis codes: instead of fitting the PIXE spectra channel-by-channel, only user-selected characteristic X-ray peak areas are fitted (avoiding the difficulties of fitting background and/or detector response functions and allowing to discard peaks which are non-apt for the fit). In order to identify the peaks and filter the background, the QXAS program \cite{web:QXAS} can be used (DataFurnace can directly read its output files) although a more convenient program is currently being developed and will be incorporated into the LibCPIXE library. Besides this detail, the treatment of the PIXE spectra in DataFurnace is identical to that done for the other techniques. See \cite{Jeynes03} for a detailed description on how the fitting process works in DataFurnace, and how the information provided by different techniques about a given sample is merged.

A typical case of ambiguity found in PIXE spectra is related to the occurrence of non-resolved overlapping peaks. This can be handled either by simply ignoring them in the fit (and relaying on other peaks for those elements) or by fitting the sum instead of the separated values. Note that in many cases, the ambiguities can be also ruled out by providing additional spectra (PIXE and/or other techniques) or by setting constrains to the fit based on known information about the sample.

\section{Examples}\label{s:examples}

A first example consists in obtaining the concentration of the elements present in a bulk stoichiometric Sb$_2$S$_3$ sample. A PIXE spectrum with 1.1 MeV H$^+$ was acquired using $\phi_{inc}=15^\circ$ and $\phi_{det}=55^\circ$, where $\phi_{inc}$ and  $\phi_{det}$ are, respectively, the incidence and detection angles relative to the surface normal. A Bayesian fit of the areas of the Sb $L_{\alpha_{1,2}}$ and S $K_{\alpha_{1,2}}$ peaks was performed (the only free parameters in this case being the bulk concentration of S and Sb as well as the beam fluence). The results, as shown in Table \ref{t:results}, reproduce the expected stoichiometric values and provide a reasonable error estimation reflecting mainly the statistical uncertainties in the experimental data.

A second example consists in the analysis of a layered Mn$_4$Ir/Si sample (a Mn$_4$Ir film on top of a Si substrate). A Bayesian fit of the Mn $K_{\alpha_{1,2}}$, Ir $L_{\alpha_{1,2}}$ and Si $K_{\alpha_{1,2}}$ lines of a 1.22MeV H$^+$ PIXE spectrum (with $\phi_{inc}=7.5^\circ$ and $\phi_{det}=77.5^\circ$) was performed leaving the thickness and composition of the film as free parameters. The results are shown in Table \ref{t:results} and are compatible with the nominal values (stoichiometric concentrations and a previous RBS thickness measurement) within the errors estimated by the Bayesian inference. In order to ascertain the accuracy of these results, we performed a simultaneous RBS+PIXE analysis using the same PIXE data as before and a 2 MeV He$^+$ RBS spectrum taken at normal incidence angle and 177{\textdegree} scattering angle. As expected, the uncertainties are greatly reduced in this case (see Table \ref{t:results}) since the RBS data introduces a tight constrain in the thickness of the layer.

\begin{figure}[hbtp]
\begin{center}
\includegraphics[width=0.9\textwidth]{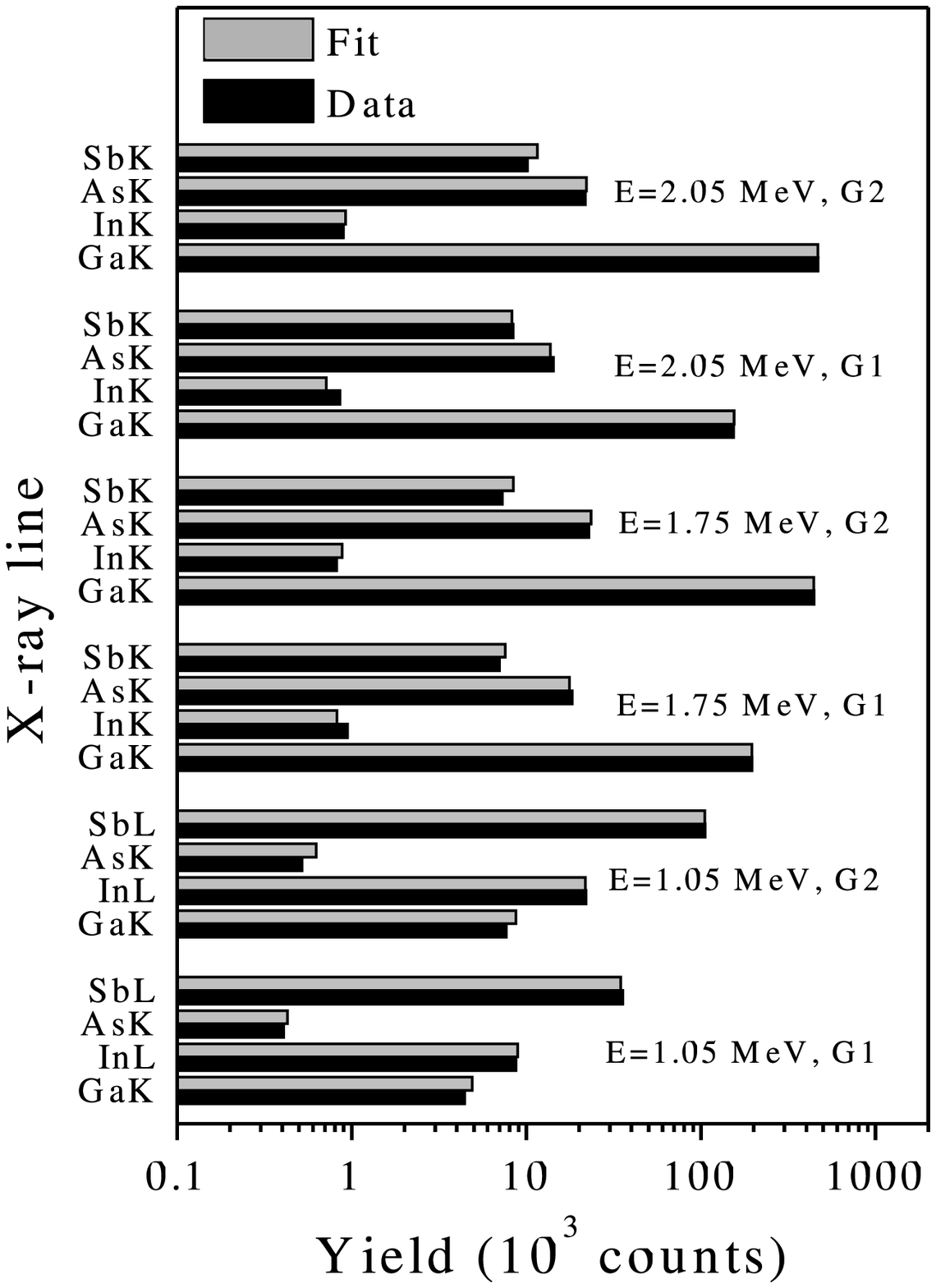}
\end{center}
\caption{Comparison between experimental and fitted main lines of the four elements in the film. $E$ is the proton beam energy. In geometry G1, $\phi_{inc}=7.5^\circ$ and $\phi_{det}=62.5^\circ$ whereas in geometry G2, $\phi_{inc}=22.5^\circ$ and $\phi_{det}=47.5^\circ$. Spectra obtained using a proton beam of 1.05 MeV were detected without any special filter between the chamber window and the Be window of the detector. The remaining spectra were measured using a 1mm thick Mylar filter in front of the x-ray detector.}
\label{f:GaInAsSb}
\end{figure}

When dealing with a more complex problem such as obtaining the concentration profiles for all elements in a GaInAsSb/GaSb sample, the information from a single PIXE spectrum (or a single RBS spectrum) is not enough to solve the ambiguities of the fit. Previous analysis of this kind of samples \cite{Corregidor05-GaInAsSb,Reis05} involved the separated analysis of RBS and PIXE spectra and a manual iteration to obtain coherent results from both techniques. The DataFurnace code now allows to perform the analysis in a much more convenient way: in the third example, the simultaneous analysis of six PIXE spectra taken with different beam energies and incidence angles proves useful for characterizing the sample. The Bayesian fit (shown in Fig. \ref{f:GaInAsSb}) was performed assuming a Ga$_{x}$In$_{1-x}$As$_{1-y}$Sb$_{y}$ ($x,y<1$) of unknown thickness and composition on top of a stoichiometric GaSb substrate. The results, shown in Table \ref{t:results}, are compatible with those of previous characterizations of the same sample (as seen in Table \ref{t:results}).

\begin{figure}[hbtp]
\begin{center}
\includegraphics[width=0.9\textwidth]{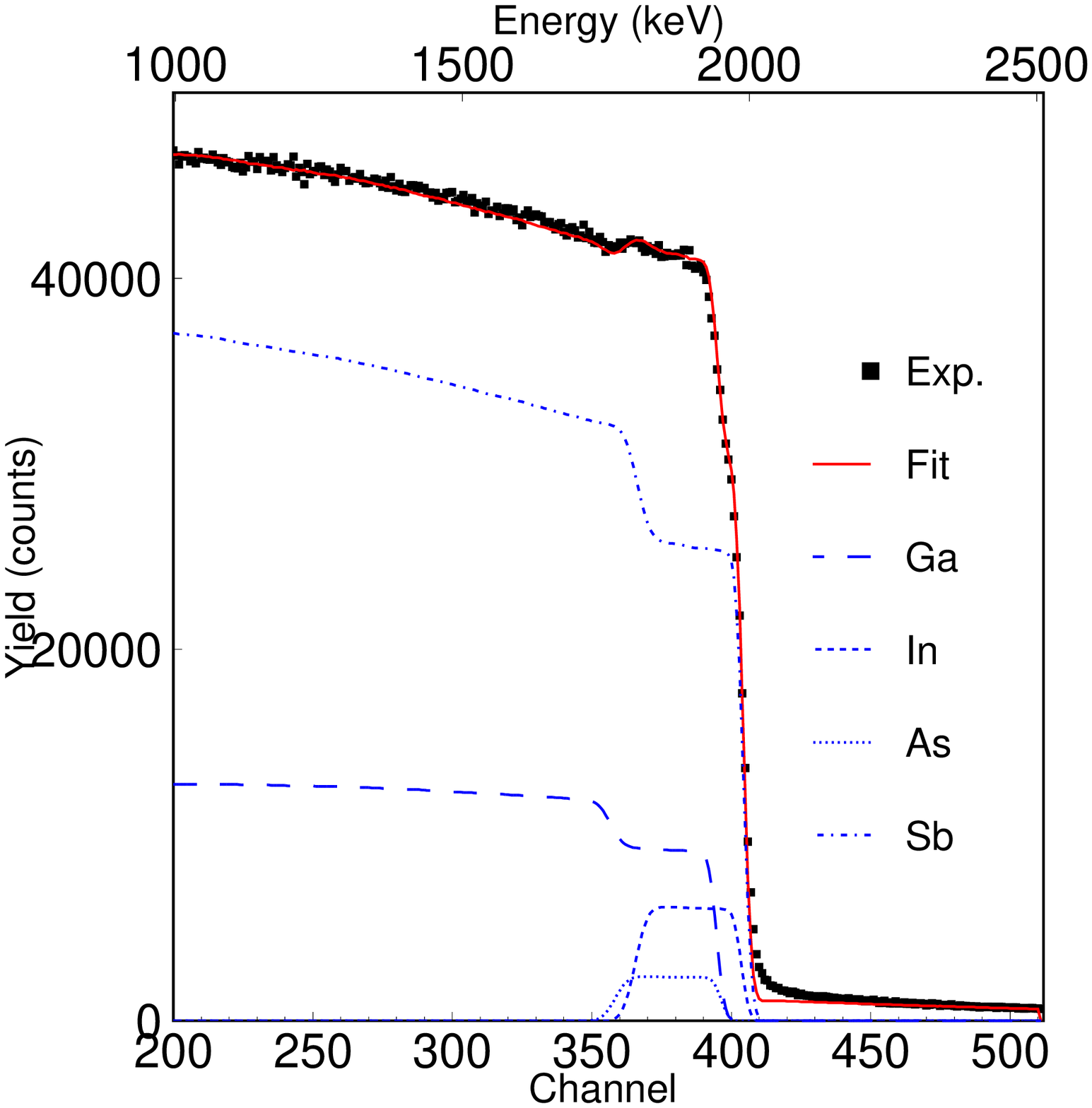}
\end{center}
\caption{Experimental RBS spectrum (squares) for 2 MeV H$^+$ on the same GaInAsSb/GaSb sample as in Fig. \ref{f:GaInAsSb}. The fit (solid line) was performed considering not only the RBS spectrum but also the data from six PIXE spectra (those of Fig. \ref{f:GaInAsSb}). Partial spectra for each element (non-solid lines) are also shown.}
\label{f:RBS}
\end{figure}

Finally, another Bayesian fit has been performed incorporating a 2 MeV H$^+$ RBS spectrum (see Fig. \ref{f:RBS}) to the mentioned set of six PIXE spectra. The results in Table \ref{t:results} show that less uncertainties are present in this case compared to the PIXE--only one due to the complementarity of the information provided by each technique: PIXE is good at determining the amount of each element but has poor depth sensitivity while RBS provides good depth resolution but cannot separate elements of similar mass (such as Ga--As or In--Sb). It is interesting to remark the fact that this analysis has been performed in one single step, providing all the experimental data to the DataFurnace code and obtaining a self-consistent result which is more reliable and considerably less time consuming than that obtained by using separated analysis codes.

\begin{table}[hbtp]
\begin{center}
\begin{tabular}{c|cc|c|cc}
Sample & \multicolumn{2}{c|}{Film Thickness ($10^{15}$at/cm$^2$)} & Element & \multicolumn{2}{c}{Atomic \%}    \\
       & fit(error) & Nominal &	 & fit(error) & Nominal \\
\hline
\hline
\rule{0pt}{3ex}
\multirow{2}{90pt}{Sb$_2$S$_3$\\(PIXE)} & \multirow{2}*{(bulk)} & \multirow{2}*{(bulk)}
	      & Sb	& 40.8(1.2)	& 40 \\
	&  &  & S	& 59.2(1.2)	& 60 \\
\hline
\multirow{3}{90pt}{Mn$_4$Ir/Si\\(PIXE)} & \multirow{3}*{551(35)} & \multirow{3}*{533$^*$}
		& Mn	& 79.1(1.2)	& 80 \\
	&  &  	& Ir	& 20.9(1.2)	& 20 \\
	&  &  	& Si$^\text{(S)}$& 100	& 100  \\
\hline
\multirow{3}{90pt}{Mn$_4$Ir/Si\\(PIXE+RBS)} & \multirow{3}*{534(3)} & \multirow{3}*{533$^*$}
		& Mn	& 79.9(0.1)	& 80 \\
	&  &  	& Ir	& 20.1(0.1)	& 20 \\
	&  &  	& Si$^\text{(S)}$& 100	& 100  \\
\hline
\multirow{6}{90pt}{GaInAsSb/GaSb\\($6\times$PIXE)} & \multirow{6}*{7917(854)} & \multirow{6}*{8442$^*$}
	 	& Ga	& 38.3(1.3)	& 38.5$^*$ \\
	&  &  	& In	& 10.7(0.9)	& 11.5$^*$ \\
	&  &  	& As	& 9.1(0.7)	& 9.0$^*$  \\
	&  &  	& Sb	& 41.9(1.3)	& 41.0$^*$  \\
	&  &  	& Ga$^\text{(S)}$& 50	& 50  \\
	&  &  	& Sb$^\text{(S)}$& 50	& 50  \\

\hline
\multirow{6}{90pt}{GaInAsSb/GaSb\\($6\times$PIXE+RBS)} & \multirow{6}*{7988(625)} & \multirow{6}*{8442$^*$}
	 	& Ga	& 40.5(0.4)	& 38.5$^*$ \\
	&  &  	& In	& 10.0(0.3)	& 11.5$^*$ \\
	&  &  	& As	& 8.8(0.3)	& 9.0$^*$  \\
	&  &  	& Sb	& 40.7(0.4)	& 41.0$^*$  \\
	&  &  	& Ga$^\text{(S)}$& 50	& 50  \\
	&  &  	& Sb$^\text{(S)}$& 50	& 50  \\

\end{tabular}
\caption{Results for the characterization of the examples of section \ref{s:examples}. The ``nominal'' values are the stoichiometric ones (when known) or had been obtained in previous independent characterizations of the same samples (marked with '$^*$'). The `(S)' superindex indicates ``Substrate'' (with known stoichiometry).}
\label{t:results}
\end{center}
\end{table}

\section{Conclusions}

We have shown that the integration of LibCPIXE into the DataFurnace code facilitates very interesting analysis modes such as the simultaneous PIXE + RBS or the multiple (differential) PIXE analysis. These combined analysis modes are, in many cases, the only way of obtaining a consistent characterization of a specimen. The use of robust fitting ---simulated annealing--- and rigorous error estimation routines ---Bayesian inference--- makes it possible to characterize arbitrary layered samples, with no limitations on the composition of each layer.

\section*{Acknowledgments}
Thanks are due to P. C. Chaves and V. Corregidor for their help with the experimental data acquisition. This work has been supported by an EU grant (HPRN-CT-2001-00199).


\end{document}